\documentstyle[epsfig]{aipproc}

\def\Msol{{\rm M_{{\odot}}}}
\newcommand{\He}{\rm{^{4}He}}

\newcommand{\Sulfur}{\rm{^{32}S}}
\newcommand{\Ar}{\rm{^{36}Ar}}
\newcommand{\Ca}{\rm{^{40}Ca}}
\newcommand{\Ti}{\rm{^{44}Ti}}

\newcommand{\Ni}{\rm{^{56}Ni}}

\newcommand{\D}{{\partial}}

\begin{document}
\title{Exploding and Non-exploding Stars:\\
       Coupling Nuclear Reaction Networks\\
       to Multidimensional Hydrodynamics}

\author{K. Kifonidis$^*$, T. Plewa$^{\dagger,*}$ and E. M\"uller$^*$}

\address{$^*$Max-Planck-Institut f\"ur Astrophysik,
         Karl-Schwarzschild-Strasse 1, D-85741 Garching, Germany \\
         $^{\dagger}$Nicolaus Copernicus Astronomical Center, Bartycka 18, 00716
         Warsaw, Poland}

\maketitle

\begin{abstract}
After decades of one-dimensional nucleosynthesis calculations, the
growth of computational resources has meanwhile reached a level, which
for the first time allows astrophysicists to consider performing
routinely realistic multidimensional nucleosynthesis calculations in
explosive and, to some extent, also in non-explosive environments.  In
the present contribution we attempt to give a short overview of the
physical and numerical problems which are encountered in these
simulations. In addition, we assess the accuracy that can be currently
achieved in the computation of nucleosynthetic yields, using
multidimensional simulations of core collapse supernovae as an
example.
\end{abstract}

\section*{Introduction}

Thermonuclear reactive flows are ubiquituous in astrophysics and
occur in non-explosive environments as, e.g., in most (hydrostatic)
stars as well as in explosive events, for which novae and supernovae
are examples. Often they provide the energy which powers stellar
outbreaks (as in the case of novae, X-ray flashes, and
thermonuclear, i.e. Type Ia, supernovae) and even for stellar
explosions where this is not the case (as e.g. in core collapse
supernovae, which are driven by neutrino heating), the strong coupling
of hydrodynamic advection and thermonuclear reactions is of utmost
importance for the nucleosynthesis which accompanies these events.
It is by a proper numerical modelling of this coupling through which a
more detailed insight into the origin of the nuclear abundances in the
solar system can be gained, which are themselves the result of a
superposition of material which has been processed in explosive and
non-explosive thermonuclear environments. By comparing the results of
numerical models with the observed solar abundance pattern, on the
other hand, one might also hope to learn more about the thermodynamic
conditions in the otherwise unaccessible nucleosynthetic sites and
events themselves.

The high precision with which nuclear abundances can be measured
nowadays poses great demands on the accuracy of the numerical models,
especially since it was convincingly demonstrated in recent years that
due to the importance of hydrodynamic instabilities, rotation, and
other effects, most of the nucleosynthetic sites do not possess
spherical symmetry.  Thus a reliable computation of the highly
non-linear interaction of hydrodynamic advection and nuclear burning
requires multidimensional numerical models.  In the following sections
we give a general overview of the methods which are currently employed
for modelling thermonuclear flows and discuss some of the problems
which are hereby encountered. Further reviews on reactive flow
modelling can be found in \cite{LeVeque98}, \cite{Mueller98} and
\cite{Oran_Boris87}.

\section*{The Governing Equations}

A rather wide range of astrophysical reactive flows, in which
relativistic effects, viscosity and magnetic fields can be neglected,
is described by the well-known (reactive) Euler equations. This
system of non-linear partial differential equations which expresses
the conservation of the total mass, momentum, total (i.e. kinetic +
internal) energy and baryons of the fluid reads
\begin{eqnarray}
   {\D \rho \over \D t } + 
   { \nabla \cdot \left(\rho \mathbf{v} \right)} &=& 0 \label{eq:cont}  \\
   {\D \rho {\mathbf{v}} \over \D t } + 
   { \nabla \cdot \left(\rho \mathbf{v} \mathbf{v} \right)} + 
   \nabla P  & = & \rho \mathbf{g} + \rho \mathbf{f}_{\rm add}  \label{eq:momentum} \\
   {\D \rho E \over \D t } + 
   { \nabla \cdot \left( \left[ \rho E + P \right] \mathbf{v} \right)}
           & = &  \rho {\mathbf{v} \cdot \mathbf{g}}  + 
           \rho \dot Q_{\rm add} +  \rho \dot Q_{\rm nuc} \label{eq:energy} \\
   {\D \rho X_i \over \D t } + 
   { \nabla \cdot \left(\rho X_i\mathbf{v} \right)} 
   &=& \rho \dot X_i  \\
   \sum_i X_i & = & 1,  \label{eq:species}
\end{eqnarray}
where $\rho$, ${\mathbf{v}}$, $E = v^{2}/2 + e$, and $P$ have their
usual meanings, $X_i$ is the mass fraction of nucleus $i$, and $\rho
\dot X_i$ as well as $\rho \dot Q_{\rm nuc}$ are source terms due to
nuclear transmutations.  If self-gravity is important, the
gravitational acceleration
\begin{equation} 
            {\mathbf{g}} = -\nabla \Phi
\end{equation}
which appears in the source terms of Eqs.~(\ref{eq:momentum}) and
(\ref{eq:energy}) and which depends on the gravitational potential, $\Phi$,
has to be obtained from a solution of the Poisson equation
\begin{equation} 
          \Delta \Phi =  4 \pi G \rho \label{eq:poisson}.
\end{equation}
In mathematical terms Eqs.~(\ref{eq:cont}--\ref{eq:poisson}) describe
a mixed initial/boundary value problem due to the hyperbolic and
elliptic nature of the Euler and Poisson equations, respectively.
Given appropriate initial and boundary conditions, an equation of
state relating $\rho$, $P$ and $e$, and the additional source terms
$\mathbf{f}_{\rm add}$ and $\dot Q_{\rm add}$, which in general will
be problem-dependent, Eqs.~(\ref{eq:cont}--\ref{eq:poisson}) can be
solved after an appropriate {\em flow representation\/} as well as a
suitable {\em numerical scheme\/} have been adopted.

\section*{Flow Representations and Numerical Schemes}

There are two primary approaches to solve the {\em homogeneous\/} part
of the system of equations ~(\ref{eq:cont}--\ref{eq:species}). In the
Eulerian framework the system of conservation laws is solved on a grid
which is {\em fixed in space\/} and the evolution of the flow is
followed by advecting the fluid through the computational cells.  The
principal assets of this method are its straightforward extension from
one to two or three spatial dimensions and the simplicity of its
implementation on serial and parallel computer architectures.  If an
appropriate shock-capturing, finite-volume numerical scheme is used,
it is equally straightforward to obtain strict numerical conservation
of all physically conserved quantities and a sharp resolution of
shocks. The major drawback is numerical diffusion.  Consider the
continuity equation (\ref{eq:cont}) in its Eulerian form, which can be
written as
\begin{equation}
  {\D \rho \over \D t } + 
  {\mathbf{v} \cdot  \nabla} \rho +  
   \rho {\nabla \cdot \mathbf{v}} = 0,
   \label{eq:cont_euler}
\end{equation}
where the second term describes advection and the third term
compression. Numerical diffusion is introduced into a numerical
solution of this equation as a result of discretization errors of the
${\mathbf{v} \cdot \nabla}$ operator. There appears to be a simple
remedy to this problem: using the comoving derivative $ {{\rm d}/{\rm
d} t} = {\D /\D t} + {\mathbf{v} \cdot \nabla}$ we can rewrite
Eq.~(\ref{eq:cont_euler}) in the frame comoving with the matter to
obtain its {\em Lagrangian\/} form 
\begin{equation}
   {{\rm d \rho} \over {\rm d} t} + \rho \nabla \cdot {\mathbf{v}} = 0 \nonumber.
\end{equation}
Note that in this frame the advection term $\mathbf{v} \cdot \nabla
\rho$ has vanished. Therefore the Lagrangian approach is (in
principle) not prone to numerical diffusion of mass (or composition).
In Lagrangian methods each cell of the numerical grid represents a
discretized fluid element which evolves subject to forces which are
due to interactions with its neighbors and the time rate of change of
the density of such a fluid element is solely determined by the
compression (or expansion) that it experiences. Density interfaces
(contact discontinuities) as well as composition discontinuities can
be easily aligned with the boundaries of grid cells and do not have to
be advected through the grid in the course of the calculation.

While this very desirable property of the Lagrangian approach has made
it the method of choice for one-dimensional nucleosynthesis
calculations, considerable difficulties are experienced when
Lagrangian schemes are applied to multidimensional flows. Shear and
vortices can severely distort a Lagrangian grid. The discrete
approximation of differential operators over such a grid results in
large errors in the numerical derivatives, and in the extreme case
that the grid lines cross (grid tangling) the calculations have to be
stopped.  Some remapping procedure to a new, more regular grid must
then be applied which unavoidably introduces numerical diffusion to
the solution.  The distortion problem can be overcome if triangular
instead of quadrilateral grids are used \cite{Oran_Boris87} or if (as
in the Smoothed Particle Hydrodynamics, or SPH approach) no grid at
all is adopted and instead the flow is sampled by a finite number of
particles.  In the former method considerable logic overhead is added
in restructuring the deformed triangular grid, while in the latter
case, due to the Monte-Carlo nature of the sampling, Poisson noise is
introduced.

Due to the aforementioned drawbacks and due to significant progress in the
development of accurate Eulerian schemes in the early 1980's, Lagrangian
methods employing quadrilateral or triangular grids have not been used
extensively in multidimensional calculations of astrophysical flows (see
\cite{Woodward76}, \cite{Pen98} as well as \cite{Oran_Boris87} and the
references therein for examples).  On the other hand, the simplicity of
SPH has made this method very popular for astrophysical (especially
cosmological) simulations. Without attempting to escalate the very
vigorous discussion, whether SPH or grid-based Eulerian schemes are to
be prefered in astrophysical calculations (see e.g. \cite{Mueller98}),
we will argue below that, due to its Monte-Carlo nature, the SPH
scheme appears to be rather unsuited for multidimensional
nucleosynthesis calculations, especially in cases where hydrodynamic
instabilities are known to be important.

\begin{figure*}[t]
\centering
\begin{tabular}{c}
\resizebox{0.65\hsize}{!}{\includegraphics{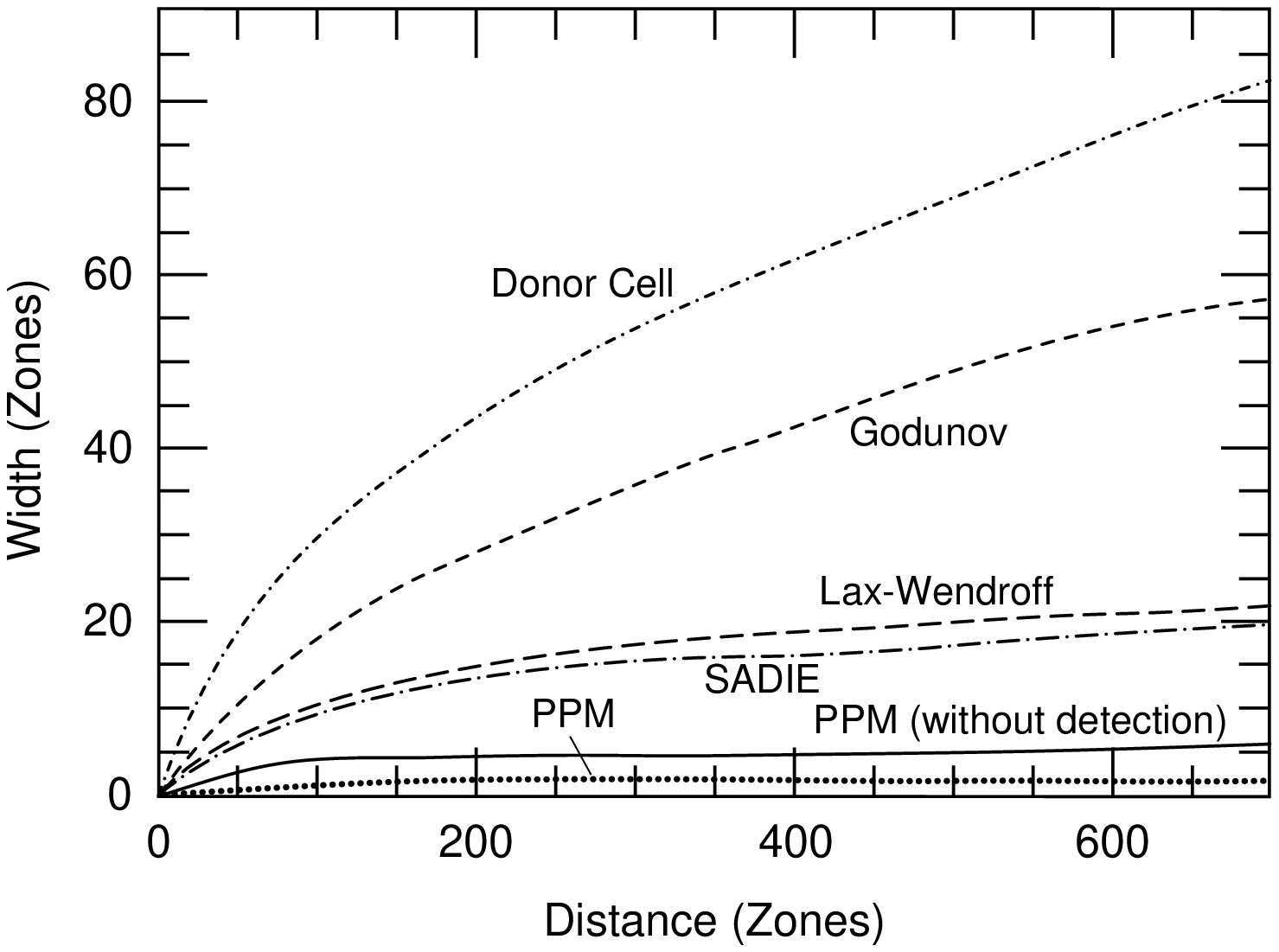}}
\end{tabular}
\caption{Comparison of the diffusivity of different advection schemes
         for the problem of the propagation of a contact discontinuity
         through an Eulerian grid. The curves give the width of the
         discontinuity (in grid zones) as a function of the number of
         zones it has propagated through the grid (adapted from
         \protect\cite{FMA89} and \protect\cite{FMA91}).}
\label{fig:fma89}
\end{figure*}

Among Eulerian schemes, the so-called shock-capturing schemes have
proven to be the most accurate ones for problems which involve
discontinuities in the flow as shock waves (see \cite{WC84} for
details). The latter are very frequently encountered in explosive
events, since in these cases the flows can attain supersonic
speeds. Shock-capturing schemes derive their accuracy from a
discretization of the hydrodynamic equations which closely mimics the
physics of compressible flows by making use of the Riemann problem,
i.e. the dissolution of an arbitrary flow discontinuity into a set of
simple waves (shocks, contact discontinuities and rarefaction waves).
Suitably constructed Riemann problems at the interfaces between
adjacent computational cells are solved within each time step, from
which the complete solution of the system of conservation laws is
constructed. This allows one to avoid the use of large amounts of
artificial viscosity in order to obtain a well-behaved numerical
scheme in the vicinity of shocks. One of the most accurate
shock-capturing schemes, which has been widely applied in
astrophysics, is the (direct Eulerian) PPM scheme of \cite{CW84}, a
second order extension of Godunov's original (and rather diffusive)
first-order shock-capturing scheme \cite{Godunov}.  In addition to its
accurate treatment of shocks PPM includes a special detection and
steepening algorithm to minimize numerical diffusion across contact
discontinuities.

The superiority of shock-capturing schemes in computing
compressible flows has been demonstrated e.g. in \cite{WC84}, and
their performance for computing reactive astrophysical flows was
studied in \cite{FMA89} and
\cite{Plewa_Mueller99}. Fig.~\ref{fig:fma89} shows a representative
result from \cite{FMA89} in which PPM was compared to a number of
older Eulerian schemes which were in wide-spread use until the mid
1980's. The figure shows the width of a contact discontinuity as a
function of the number of zones that it has travelled across a
numerical grid. Most Eulerian schemes tend to smear such fluid (and
also composition) interfaces without limit, i.e. the width of the
``discontinuity'' tends to grow with time. Of all the schemes
investigated, only PPM maintained a sharp resolution of the interface
within two zones. Still however, numerical diffusion cannot be
completely avoided in Eulerian calculations and its minimization
necessitates an adequate spatial resolution in addition to an excellent
advection scheme. This has led to the development of adaptive mesh
refinement methods \cite{BC89}, which concentrate the computational effort in
critical regions of the flow and thereby often allow for substantial
savings in computer time.

\section*{Additional Physics}

While the numerical problems encountered in solving the homogeneous
part of the Euler equations are difficult to overcome, they represent
only a part of the computational difficulties for a realistic
simulation. The source terms, which are usually taken into account
using the operator splitting technique \cite{LeVeque98}, often require
much more computer time than the solution of the hydrodynamic
equations themselves. This holds, e.g. if large nuclear networks need
to be evolved with the hydrodynamics or transport processes need to be
taken into account (as e.g. neutrino transport in core collapse
supernovae, see \cite{Rampp_Janka00} and the references therein and
A. Burrows, this volume). In some cases even phenomenological
(sub-grid) models might have to be introduced.  This is e.g.\ the case
for turbulent combustion in thermonuclear supernovae, where a white
dwarf is incinerated by a deflagration front whose propagation speed
is impossible to compute in a direct simulation since this would
require a resolution of the turbulent energy cascade down to the
dissipation length scale \cite{Reinecke+99}.  Exacerbating the
situation is the fact that stellar models, which serve as initial data
for supernova simulations, might be affected by considerable
uncertainties. In the absence of computational schemes and resources
which allow for a consistent multidimensional treatment of stellar
convection and rotation over stellar evolutionary time scales, one is
forced to describe these phenomena by one-dimensional appoximations
(see the contribution of N. Langer, this volume). Finally,
uncertainties in nuclear reaction rates enter the calculations.

It is apparent that progress in only a {\em single\/} of the
involved fields is {\em not\/} going to improve the accuracy of the
desired nucleosynthetic yields considerably. In fact, a concerted
effort in all areas appears to be required, since, as we will show in
our example below, the different effects can conspire in falsifying
the nucleosynthetic yields.

\section*{Core Collapse Supernovae: A case study}

Nucleosynthesis in core collapse supernovae is a good example for
illustrating the aforementioned problems and we will start with a
discussion of numerical diffusion using results of simple
one-dimensional calculations.  We subsequently address the
complications introduced by convection in multidimensional
calculations as well as by ``additional physics'', i.e. neutronization
due to neutrino matter interactions.  Finally we show how a
multidimensional numerical failure, the so-called
``odd-even-decoupling'' phenomenon, an instability which appears to
plague most shock-capturing schemes and whose effects have not yet been
discussed extensively in the numerical astrophysics literature, can
enhance neutronization by strengthening hydrodynamic convection and
affect the nucleosythetic yields in multidimensional simulations.

\subsection*{Nucleosynthesis in a 15\,$\rm M_{\odot}$ star (1D)}

\begin{figure}[t]
\begin{center}
\begin{tabular}{cc}
{\includegraphics[bb=60 110 480 720, width=0.5\columnwidth]{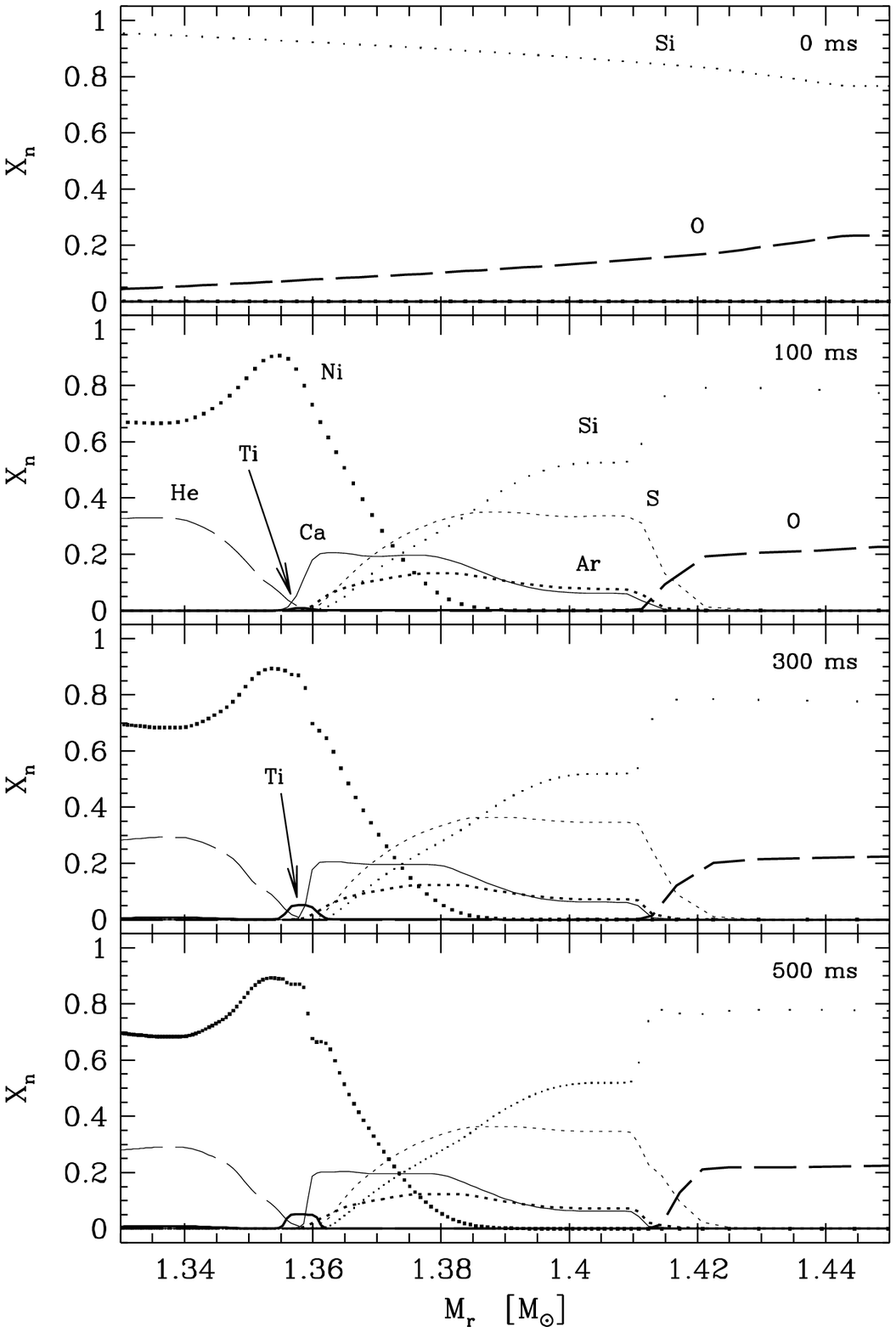}} &
{\includegraphics[bb=70 110 490 720, width=0.5\columnwidth]{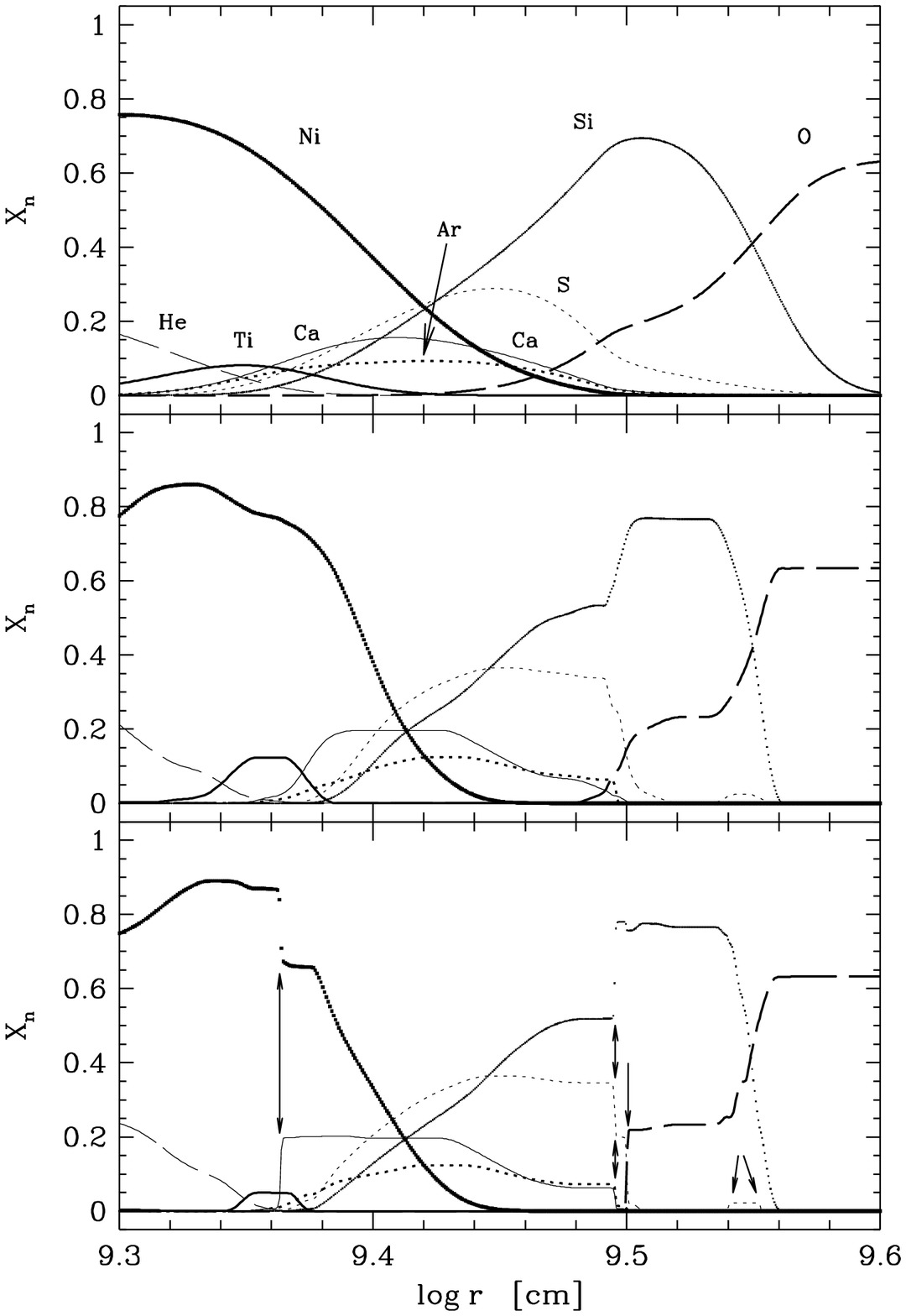}} \\
\end{tabular}
\end{center}
\caption{Left: Eulerian PPM calculation of explosive nucleosynthesis in the
         presupernova model of \protect\cite{WPE88} using an $\alpha$-nucleus
         network and the Consistent Multifluid Advection scheme (CMA)
         (from \protect\cite{Plewa_Mueller99}).  Right: Comparison of Eulerian
         PPM results using: 1st order advection for nuclear species
         (top), the FMA advection scheme for multifluid flows of
         \protect\cite{FMA89} (middle), and the CMA scheme of
         \protect\cite{Plewa_Mueller99} (bottom). Note the
         decreasing amount of diffusion and the sharp interfaces
         obtained with CMA (from \protect\cite{Plewa_Mueller99}).
         }
\label{fig:cma_plots}
\end{figure}

In core collapse supernovae nucleosynthesis is triggered by a shock
wave which forms after the collapse of the iron core of a massive star
has proceeded to supranuclear densities. The shock, while initially
powerful, stalls after a few milliseconds due to the energy losses
from which it suffers while propagating through the outer iron core,
but eventually ejects the outer stellar layers if heating by neutrinos
from the collapsed core is able to overcompensate for the energy
losses.  In most nucleosynthesis calculations, however, these
processes are not modelled in detail and instead a shock is initiated
by simply depositing the typical observed supernova energy of $\sim
10^{51}$\,erg near the center of a presupernova model.

\begin{figure*}[t]
\begin{center}
\begin{tabular}{c}
\resizebox{0.7\hsize}{!}{\includegraphics{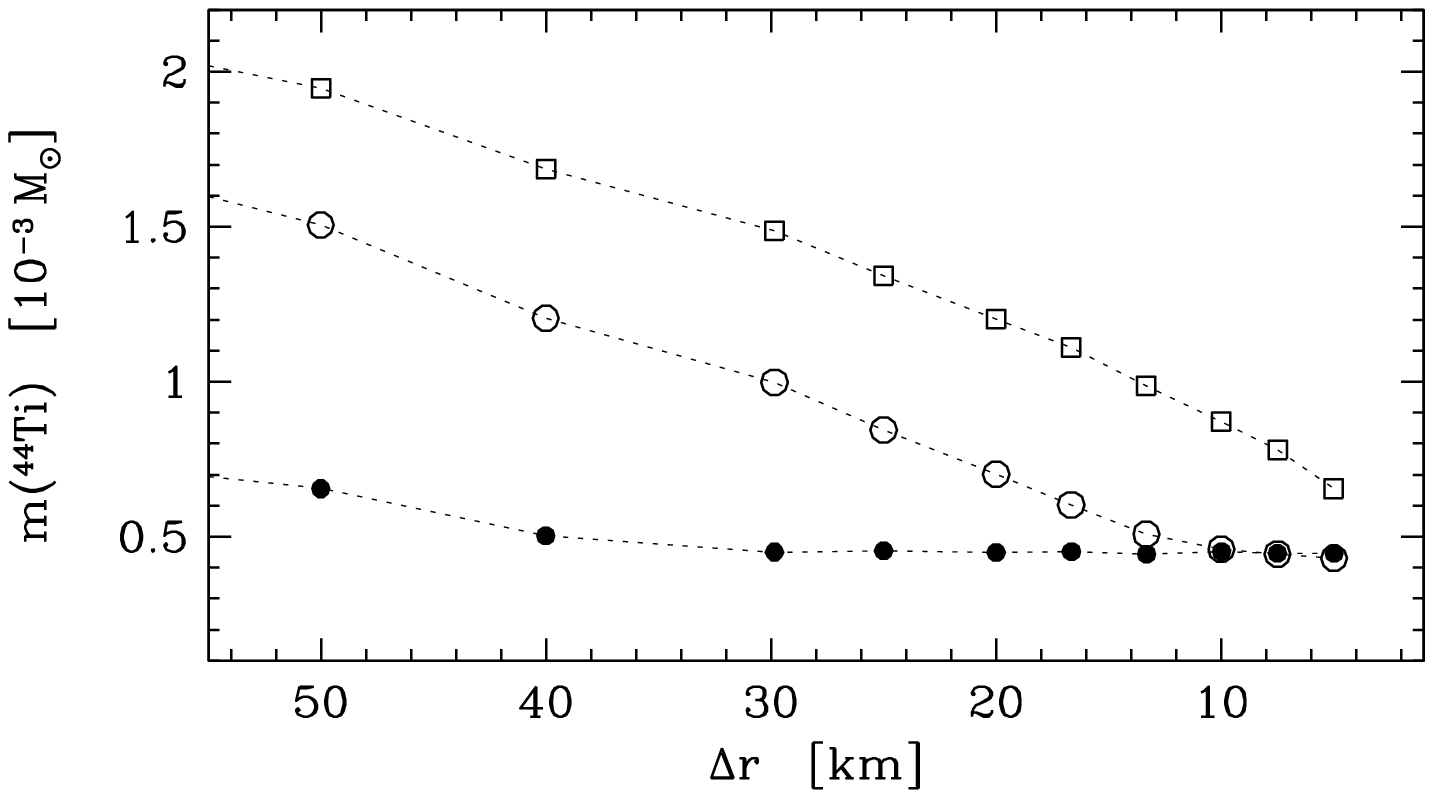}}
\end{tabular}
\end{center}
\caption{Final $\rm ^{44}Ti$ yield of the one-dimensional calculations
         shown in Fig.~\ref{fig:cma_plots} as a function of radial
         resolution and different multifluid advection schemes.  Top
         curve: 1st order species advection. Middle: FMA. Bottom: CMA
         (from \protect\cite{Plewa_Mueller99}).}
\label{fig:ti44_yield}
\end{figure*}

\begin{figure*}[t]
\centering
\begin{tabular}{c}
\resizebox{0.8\hsize}{!}{\includegraphics{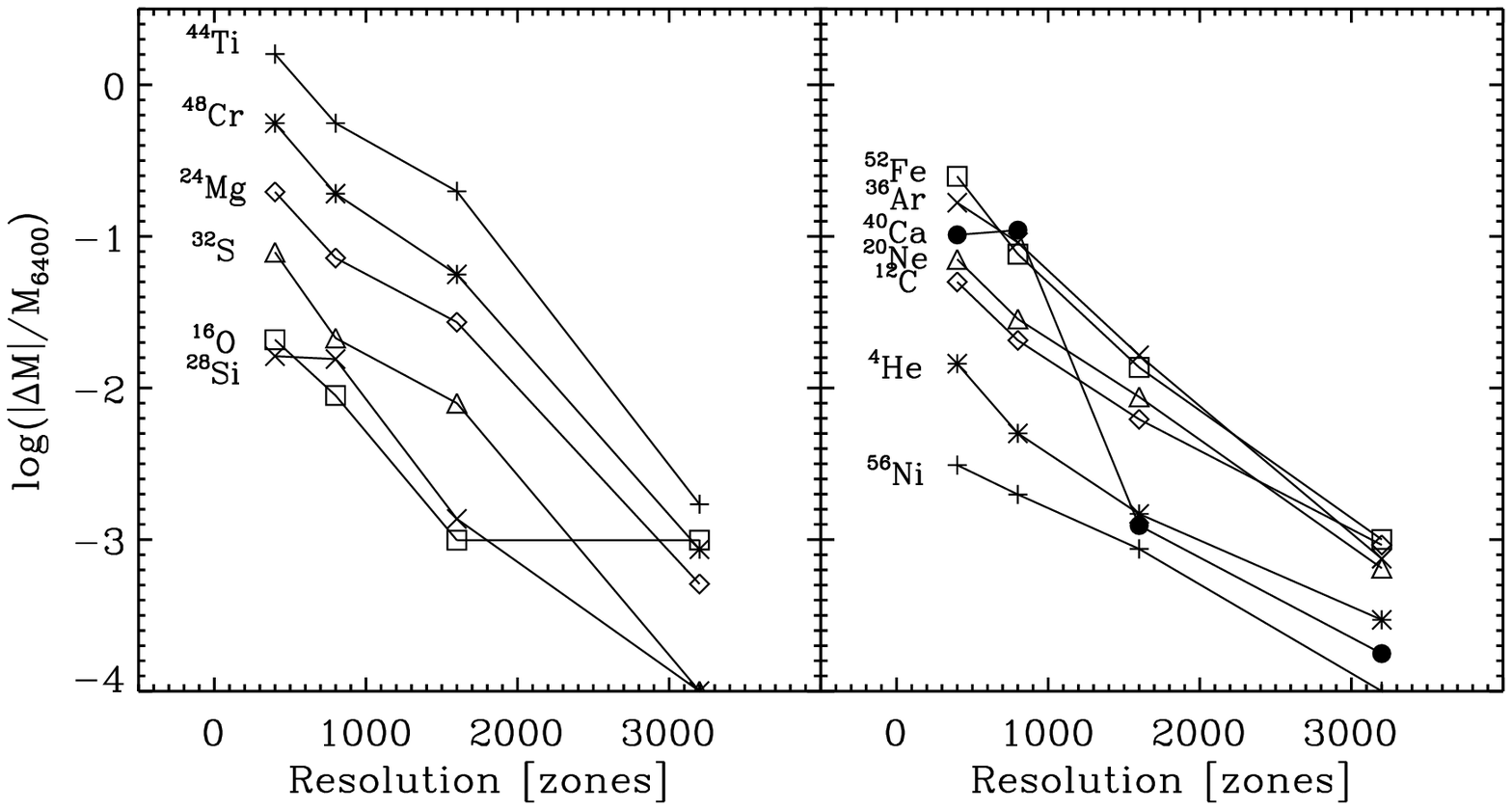}}
\end{tabular}
\caption{Accuracy of various elemental yields from one-dimensional
         shock-revival and explosive nucleosynthesis calculations in
         the post-bounce model of \protect\cite{Bruenn93} (see also
         \protect\cite{WPE88} for the presupernova model). The
         logarithm of the deviations of the elemental yields of
         various $\alpha$-nuclei as compared to an essentially
         converged 6400 zone calculation ($\Delta r \leq 5$\,km) is
         displayed as a function of the radial resolution (in grid
         zones)(from \protect\cite{Kifonidis00}).}
\label{fig:yields}
\end{figure*}

In the left panels of Fig.~\ref{fig:cma_plots} we show snapshots of
the mass fractions from the first 500\,ms of such a calculation,
focusing on the silicon-rich layers just outside the iron core in
which explosive nucleosynthesis takes place. Only 100\,ms after the
start of the calculations explosive silicon burning has frozen out and
has left behind a significant abundance of $\Ni$ as well as $\Ca$,
$\Ar$ and $\Sulfur$. Particularly noteworthy for the following
discussion is the nucleus $\Ti$. From one-dimensional Lagrangian
nucleosynthesis calculations \cite{TNH96}, \cite{Woosley_Weaver95} it
is known that this isotope should be primarily synthesized in the
innermost stellar layers which experience an $\alpha$-rich freezeout.
However, in the present Eulerian calculation a significant abundance
of $\Ti$ has formed in zones with mass coordinates around
1.36\,$\Msol$ (marked with arrows in the left panels of
Fig.~\ref{fig:cma_plots}), i.e. at the interface of the regions
enriched in $\He$ and $\Ca$.  This $\Ti$ ``bump'' results from the
reaction $\Ca(\alpha,\gamma)\Ti$ and the amount of $\Ti$ thereby
produced is very sensitive to numerical diffusion in this region.
Consequently, the strength of the $\Ti$ ``bump'' varies with the
diffusivity of the numerical scheme which is used to advect the
nuclear species. This is illustrated in the right panels of
Fig.~\ref{fig:cma_plots}.  The top right panel depicts results from a
calculation with Godunov's first-order scheme.  In this case, all mass
fraction profiles are heavily smeared due to strong diffusion, as can
be seen by a comparison to the middle right and bottom right panels
which show results that were obtained with the FMA and CMA advection
schemes of \cite{FMA89} and \cite{Plewa_Mueller99}, respectively. Of
all these schemes, CMA is the least diffusive since it is the only
method which includes a detection and steepening algorithm for
composition interfaces which was derived from PPM's original detection
and steepening algorithm for contact discontinuities.  Note the size
of the $\Ti$ bump for the three different runs. The more diffusive
schemes produce much more $\Ti$. This is also illustrated in
Fig.~\ref{fig:ti44_yield} which summarizes how the $\Ti$ yield depends
on the adopted advection scheme and the spatial resolution.  While the
CMA results (bottom curve) are already converged for a resolution of
$\Delta r = 40$\,km, FMA (middle curve) needs a resolution of about
$10$\,km.  The first-order scheme (top curve) would need much finer
zoning than $\Delta r = 5$\,km to yield results of comparable
quality. Note also that, if a diffusive advection scheme and coarse
resolution are used, the errors might be as large as a factor of four!

It should be pointed out, however, that $\Ti$ is a somewhat extreme
(though very important) example. The (relative) errors due to
numerical diffusion are usually smaller for the more abundant nuclei.
This is illustrated in Fig.~\ref{fig:yields} which shows the
dependence of the yields of different $\alpha$-nuclei on resolution in
a 1D calculation from \cite{Kifonidis00}, in which no ad hoc energy
deposition was adopted, but where the shock revival phase was followed
in detail by including the effects of neutrino heating from a central
light bulb neutrino source \cite{JM96}.  It can be seen from this
figure that individual elemental yields are more accurate than the
$\Ti$ yield by more than an order of magnitude.  However, if one is
aiming at a numerical accuracy of about 1\% for all yields, about 3000
radial zones are required for this calculation. This amount of spatial
resolution makes accurate multidimensional simulations very expensive.

\subsection*{Convection and $\rm ^{56}Ni$ synthesis 
             in core collapse supernovae}

\begin{figure*}[t]
\centering
\begin{tabular}{c}
\resizebox{0.88\hsize}{!}{\includegraphics{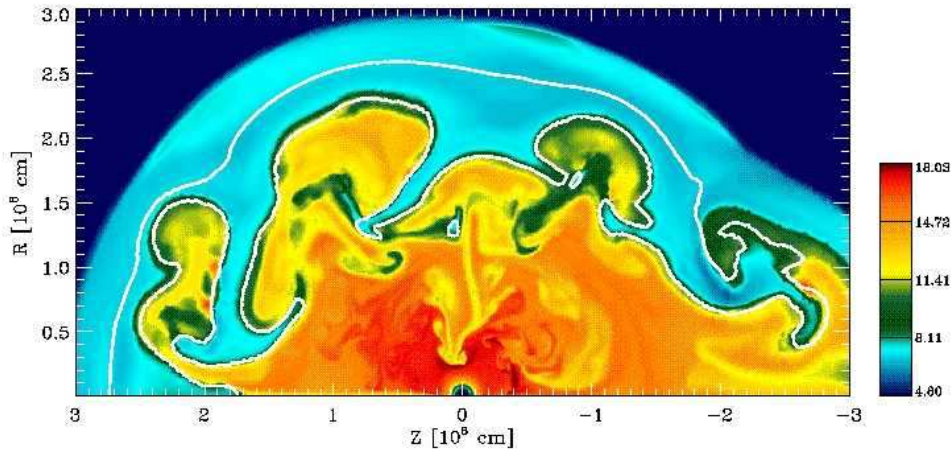}}
\end{tabular}
\caption{Distribution of the entropy (in units of $k_B/{\rm nucl.}$)
         320\,ms after core bounce in a two-dimensional core collapse
         supernova simulation of a 15\,$\Msol$ star.  The white
         contour line encloses the region in which the $\rm ^{56}Ni$
         mass fraction exceeds 20\% (from \protect\cite{Kifonidis00}).}
\label{fig:entropy_245}
\end{figure*}

Neutrino matter interactions play a crucial role in the explosion of
core collapse supernovae. They heat the matter behind the stalling
shock and thereby trigger the explosion.  On the other hand they
determine the electron fraction per baryon, $Y_{\rm e}$, (or
equivalently the ratio of protons to neutrons) in the ejecta and thus
influence the nucleosynthetic yields. If the $Y_{\rm e}$ value of
material that has been photodissociated by the shock is significantly
reduced below 0.5 by neutrino/matter interactions, this matter will
recombine mainly to neutron-rich nuclei after expanding and cooling.
In that case nuclei with $Z=N$ like $\Ni$ will not form in the ejecta.
Fig.~\ref{fig:entropy_245} which shows results of a two-dimensional
simulation of shock revival illustrates this effect. In this
calculation the luminosities of $\nu_e$ and $\bar \nu_e$ were such
that $\bar \nu_e$ absorption on protons was favored against $\nu_e$
absorption on neutrons and thus the heated gas was neutronized
\cite{JM96}. The sharp division between this material which is visible
in the bubbles behind the shock in Fig.~\ref{fig:entropy_245} and the
lepton-rich material farther out which has formed $\Ni$ (the region
enclosed by the white contour line in Fig.~\ref{fig:entropy_245}) is
clearly visible. The negative entropy gradient that the heating has
imprinted on the layers between the radius of maximum neutrino heating
(close to the center) and the shock farther out, has also led to
strong convective motions: bubbles of heated deleptonized gas rise
toward the shock while lower entropy flux tubes transport lepton-rich
matter to deeper layers where it interacts with the neutrino fluxes
much more efficiently.  This interplay of convection and
deleptonization is crucial for the $\Ni$ yield. The shock is only able
to heat a certain amount of lepton-rich material to temperatures in
excess of the $5\times10^{9}$\,K which are required for $\Ni$
synthesis.  If convection is strong enough to advect significant
amounts of this matter close to the neutron star, where the gas will
experience deleptonization, the $\Ni$ yield will be lower than in a
model with no or weak convection.

\subsection*{Multidimensional numerical failures: Odd-even decoupling}

\begin{figure*}[t]
\centering
\begin{tabular}{c}
\resizebox{0.88\hsize}{!}{\includegraphics{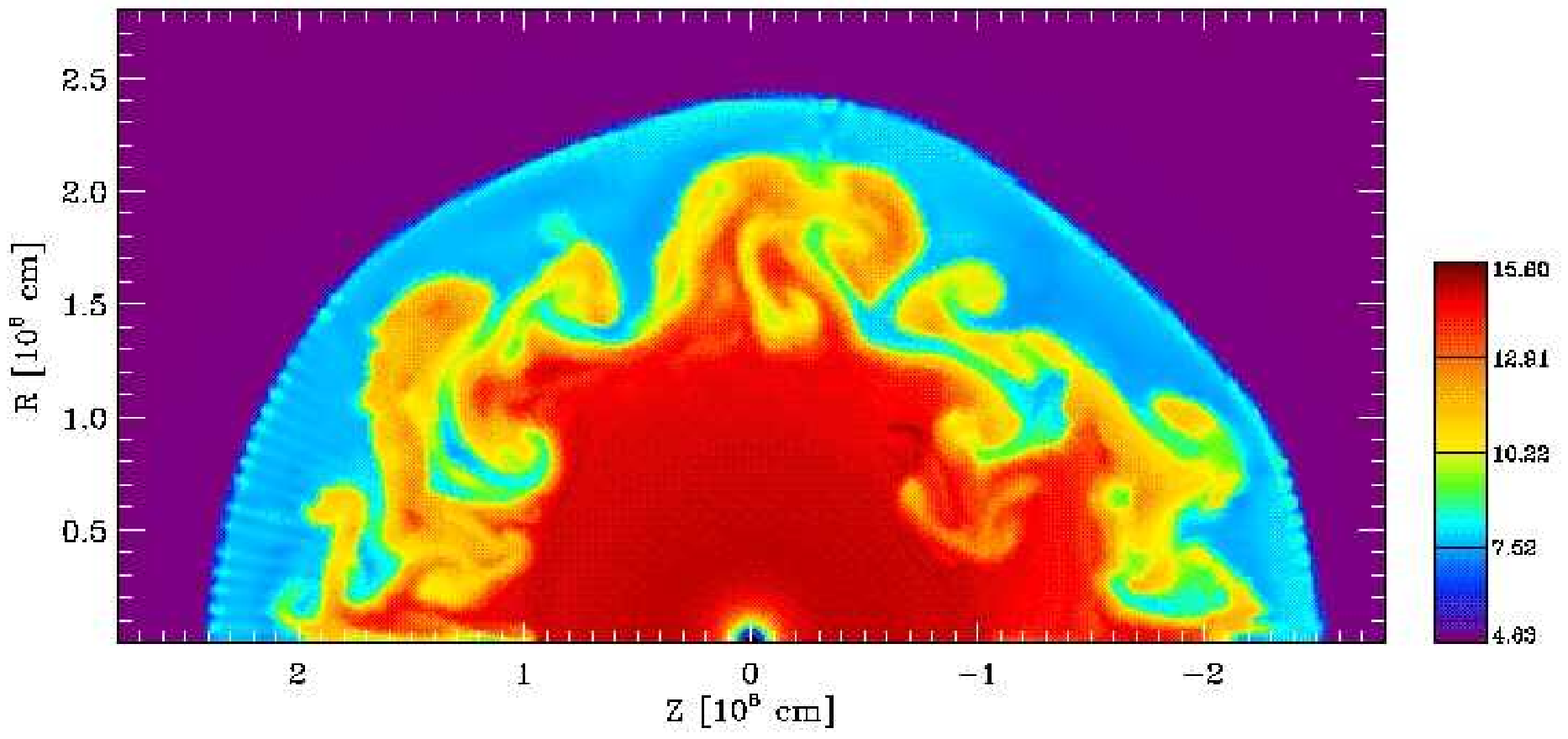}} \\
\resizebox{0.88\hsize}{!}{\includegraphics{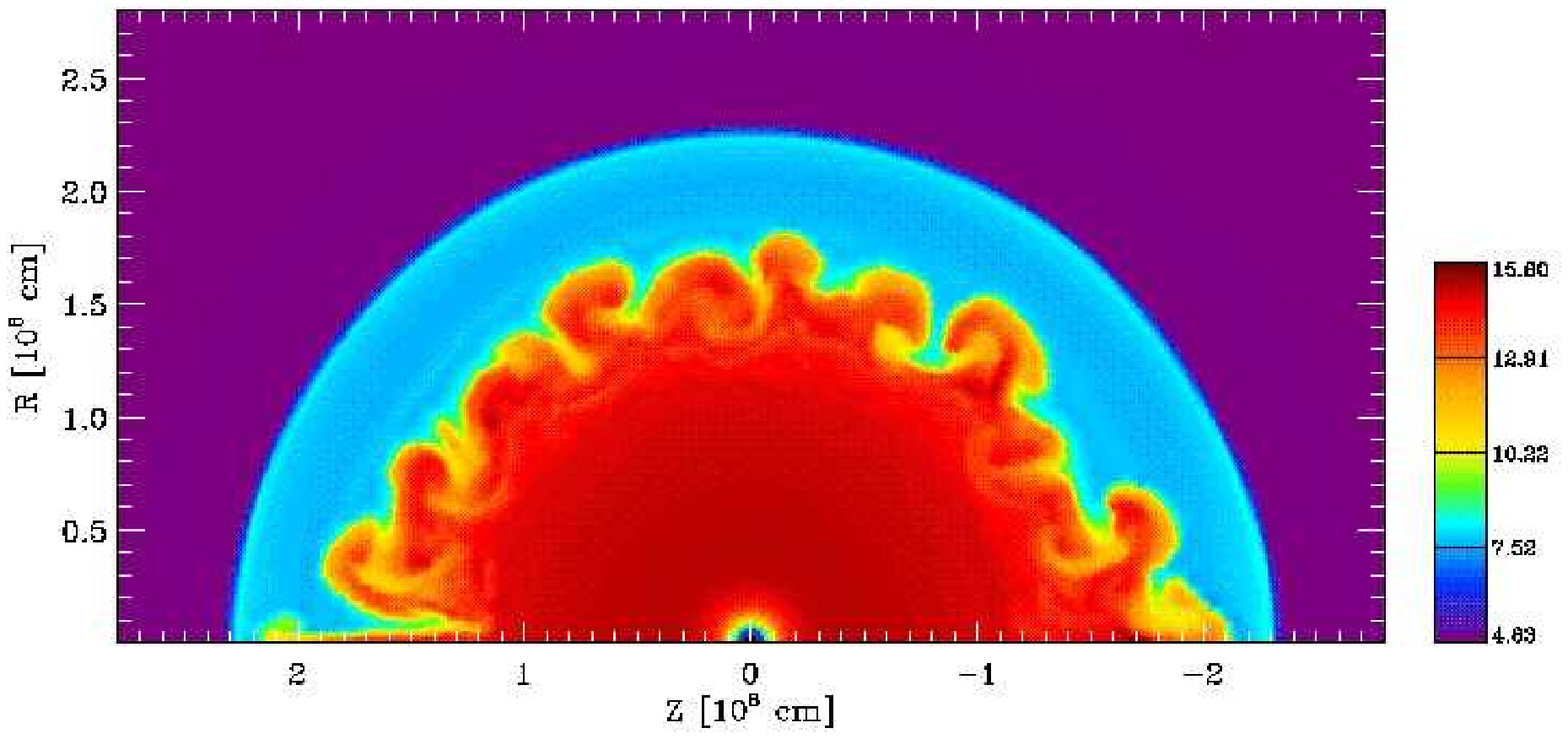}}
\end{tabular}
\caption{Top: Entropy distribution 208\,ms after core bounce (in units
         of $k_B/{\rm nucl.}$) in a two-dimensional supernova model
         showing odd-even decoupling.  Bottom: Entropy distribution  
         for an equivalent calculation in which odd-even decoupling has
         been suppressed (from \protect\cite{Kifonidis00}).}
\label{fig:entropy_odd-even}
\end{figure*}

Quirk \cite{Quirk94} has reported on a subtle flaw in a number of
shock capturing schemes which becomes evident when calculating
multidimensional flows with strong, grid-aligned shocks.  He has
dubbed this failure the ``odd-even decoupling'' phenomenon.  The
problem shows up only if a sufficiently strong shock is either fully
or nearly aligned with one of the coordinate directions of the grid,
and if, in addition, the flow is slightly perturbed. This can be due
to either perturbations intentionally introduced in order to study
\emph{physical} instabilities, as it is done in all studies of
convection in supernovae, or due to perturbations caused by other flow
features. Many Riemann solvers show the tendency to allow these
perturbations to grow without limit along the shock surface, thus
triggering a strong rippling of the shock front as well as the post
shock state.  In supernova simulations these perturbations, whose
amplitudes can exceed those of the seed perturbations by several
orders of magnitude, enhance the growth of hydrodynamic
instabilities. In case of neutrino driven convection they lead to
large-scale overturn and angular wavelengths of convective bubbles
which are significantly larger than in a ``clean'' calculation.  This
artificial enhancement of convection is demonstrated in
Fig.~\ref{fig:entropy_odd-even} where the entropy distribution of a
simulation exhibiting odd-even decoupling (top panel) is compared to
one in which the numerical failure has been cured (bottom panel).  A
modification of PPM's original dissipation algorithms \cite{CW84} was
necessary for this purpose.  Alternatively the hybrid Riemann solver
method of \cite{Quirk94} might be used.  Note that the calculations
have been carried out in spherical coordinates $(r,\theta)$ so that
the (initially spherical) shock wave was fully aligned with the grid.
However, cylindrical coordinates which have been used for plotting are
indicated in the figures.  The difference in the final $\Ni$ yield for
these two simulations was about 40\%, the calculation not exhibiting
odd-even decoupling showing the larger yield, as expected from our
discussion in the previous section. This demonstrates that due to the
strong coupling of neutrino physics and convection, the $\Ni$ yield in
multidimensional calculations is much more difficult to calculate
correctly than in one spatial dimension. Therefore the error of at
most a few percent which can be deduced for the latter case from
Fig.~\ref{fig:yields} can be deceptive. The results shown in
Fig.~\ref{fig:entropy_odd-even} also suggest that numerical noise,
whatever its origin is, leads to a grossly overestimated efficiency of
convection. Thus, schemes which are known to suffer from this problem
(like SPH) do not appear to be suited for nucleosynthesis calculations
in core collapse supernovae.

\section*{Conclusions}

Realistic nucleosynthesis calculations in astrophysical contexts
represent a challenge in many respects. The difficulties involve the
numerical treatment of multidimensional hydrodynamic advection,
complex physics in addition to hydrodynamics and burning, disparate
length and time scales, realistic initial conditions and uncertainties
in reaction rates. It is our conviction, that substantial efforts are
required in {\em each\/} of these fields in order to obtain reliable
yields in multidimensional nucleosynthesis calculations.

\end{document}